# Polarization dynamics, stability and tunability of a dual-comb polarization-multiplexing ring-cavity fiber laser

**ALBERTO RODRIGUEZ CUEVAS,[1] HANI J. KBASHI,[1] DMITRII STOLIAROV, [1] SERGEY SERGEYEV,[1]*** 

[1] *Aston Institute of Photonic Technologies, College of Engineering and Physical Sciences, Aston University B4 7ET Birmingham, United Kingdom*

**s.sergeyev@aston.ac.uk*

**Abstract:** In this paper, we demonstrate the polarization-multiplexed system capable of generating two stable optical frequency combs with tunable frequency differences in the range from 100 to 250 Hz and an extinction ratio of 16.5 dBm. Also, the polarization dynamics of a dual-frequency comb generated from a single mode-locked Er-doped fiber laser are experimentally studied. The obtained results will extend the application to areas such as polarization spectroscopy and dual-comb-based polarimetry.

## 1. Introduction

Dual frequency combs (DFC) consisting of two pulse trains with slightly different repetition rates have been used in several applications over the last two decades. These applications include distance ranging (LIDARs) [1,2], structural health monitoring [3,4], gas sensing [5–7], optical communications [8], and calibration of instrumentation for astronomy [9], among others. To the date, there are various ways to generate dual optical frequency combs both in fiber lasers and solid-state lasers. Nonetheless, the traditional implementations based on two synchronized ultrafast lasers require addressing several technical challenges, such as the complexity of the phase-locking between both lasers. Therefore, most research efforts have been recently re-directed towards simpler systems based on a single laser.

Combs have two degrees of freedom: repetition rate $f_{rep}$ and carrier-envelope offset frequency $f_{CEO}$. Though individual $f_{rep}$ ($f_{CEO}$) of a free-running dual-comb in a single mode-locked laser drifts over time, the offset between the two combs $\Delta f_{rep}$ ($\Delta f_{CEO}$) tends to show superior long-term stability. This type of stability reflects intrinsic mutual coherence between the two pulse trains which is essential for dual-comb applications [10, 11]. Thus, a single cavity-based dual-comb is a cost-effective solution without complex $f_{rep}$ and $f_{CEO}$ stabilization based on Pound-Drever-Hall (PDH) and phase-locked loop techniques [11].

The dual-comb single-cavity laser design approaches include wavelength multiplexing, polarization multiplexing, circulation direction multiplexing, cavity space multiplexing, and extra-cavity fiber delay methods [11]. The wavelength-multiplexing explores two trains of pulses with different central frequencies traveling simultaneously and in the same direction within the cavity [10,11]. In polarization multiplexing, two combs coexist in the same cavity circulating in the same direction but with two orthogonal states of polarization (SOPs). Given the adjustment in the in-cavity birefringence, a difference in the repetition rate of both combs appears [12]. In the circulation-direction multiplexing, two trains of pulses travel in opposite directions of the same cavity. These pulses interact slightly differently with the cavity, generating combs with two different repetition rates [13,14]. In the cavity-space multiplexing, a segment of the laser cavity is split into two segments with two different components, such as two different isolators or two different gain media for creating two different trains of pulses [15]. Finally, the so-called extra-cavity methods comprise a single frequency-comb laser source and, for example, the Michelson interferometer or other methods that produce two combs from the original output [11].

To date, various studies have been done where polarization multiplexing has been used to generate two combs in one laser [10, 11, 16]. By introducing a piece of highly birefringence material in the medium, usually a segment of polarization-maintaining (PM) fiber, the refractive index of the cavity is diverted into two axes, the slow and fast axis of the PM fiber. Therefore, the trains of pulses with two orthogonal SOPs interact differently with the birefringent media in the cavity, for example, PM fiber, which results in different repetition rates. Unlike dual-wavelength lasers, the dual-polarization comb takes the form of two pulse trains with an overlapping spectrum and supports intrinsic spectral coherence. To the best of our knowledge, there has not been any prior study evaluating the stability and dynamics of the orthogonal SOPs in similar laser cavities. Nonetheless, the insight into SOPs' dynamics could extend the range of dual-comb laser applications to ellipsometry and environmental monitoring [17, 18]. To address this gap, in this paper, we study experimentally the stability, tunability, and polarization dynamics of this dual-comb laser based on the polarization-multiplexing technique.

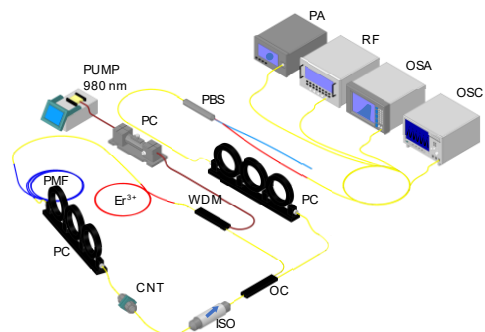

Fig. 1. 3D representation of the laser cavity consisting of the following parts: isolator (ISO), optical coupler 90 / 10% (OC), wavelength-division multiplexing (WDM), polarization controller (PC), saturable absorber based on a carbon nanotubes (CNT), segment of polarization maintaining (PM) fiber, fiber segment of Erbium-doped fiber (Er3+), and a continuous wave laser diode at 980nm (PUMP). Outside of the cavity there is a polarization controller (PC) and a polarization beam splitter (PBS) in combination with the following measuring devices: polarimeter (PA), radiofrequency spectrum analyzer (RF), optical spectrum analyzer (OSA) and oscilloscope (OSC).

## 2. Set up design

A schematic representation of the polarization-multiplexed system based on a fiber ring laser is displayed in Figure 1. 0.85-meter-long Er-doped fiber (Liekki Er80-8/125) with group velocity dispersion (GVD) parameter of - 0.017 $ps^2$ is pumped using a 980 nm laser diode through a 980/1550 nm WDM and an inline polarization controller to control SOP of the pump. A one-meter-long PM-1550 fiber with GVD of – 0.023 $ps^2$, a numerical aperture of 0.125 and core/cladding diameters of 8.5/125 µm has also been used inside the cavity to generate polarization multiplexing mode-locked regime, hence, to generate two trains of pulses in the same laser cavity. This polarization multiplexing can only be achieved by adjusting the polarization states inside the cavity with a polarization controller. Thus, a 3-paddle polarization controller comes after the PM fiber. A 51 dB dual-stage polarization-independent optical isolator (Thorlabs IOT-H-1550A) is inserted in the laser cavity to ensure single direction propagating. A film-type homemade single-wall CNT absorber has been inserted between fiber connectors as a transmission-type mode-locker. Finally, an output optical coupler redirects 10% of light outside the cavity. In total, the length of the cavity is 17.3 m, while the roundtrip time of a pulse is 83 ns (around repetition rate $f_{rep}$ =12 MHz). In contrast to the majority Er-doped single-mode fibers, Er80-8/125 has negative GVD, consequently, we got an all-anomalous net cavity dispersion at 1550 nm with the parameter of - 0.38 $ps^2$.

For the polarization-multiplexed comb separation, the polarization controller and polarization beam-splitter were connected to the laser output. The detection and measuring systems include a photodetector (InGaAsUDP-15-IR-2_FC) with a bandwidth of 17 GHz connected to a 2.5 GHz sampling oscilloscope (Tektronix DPO7254). A fast polarimeter (Novoptel PM1000-XL-FA-N20 D) with a sampling frequency of 100 MS/s to measure the normalized Stokes parameters s1, s2, s3, and degree of polarization (DOP). An optical spectrum analyzer (Yokogawa AQ6317B) with a maximum resolution of 20 pm, and a radio frequency spectrum analyzer (FSV Rohde Schwarz) have also been used to measure the optical and electrical spectrums respectively.

The PM fiber and the in-cavity polarization controller act as an optical filter inside the laser cavity by adjusting a high level of birefringence in the media and causing the formation of the two combs. Some authors have already demonstrated the ability of this configuration to generate tunable dual comb, they have used a different length of PM and dispersion compensated fiber segments and other components [12]. Additionally, a similar configuration was used to generate a phase-stable dual wavelength regime [19]. Nonetheless, in this last work they used a 10-meters-long high birefringence fiber and did not used a saturable absorber.

## 3. Results and discussion

Using the setup described in the previous section, we demonstrated dual-comb emergence along with its tunability, long-term stability, and polarization dynamics. Fig.2 (a) presents the results of frequency difference tuning within a range from 110 to 250 Hz. With the correct polarization state and curvature reduction of the PM fiber, the frequency difference can go as high as 250 Hz or as low as 5 Hz. Stable dual-combs were generated at a frequency difference of more than 100 Hz. If the frequency difference is lower the regime tends to collapse. That is similar to the frequency locking phenomena observed for the coupled oscillators [20]. On the other hand, multipulsing was observed quite often when the dual-comb regime was about to be obtained. The stable dual-comb regimes were observed with phase differences in the range of 110-250 Hz. The tunability in the frequency difference was achieved by precisely adjusting the polarization controller within the cavity. It is not possible to transit from one frequency difference to another more than 10 Hz. Usually, the dual comb regime disappears and later happens again with another frequency difference. Given that the controllable birefringence in the cavity modifies the optical spectral transmission as well [11], the spectra for the frequency differences are changing as shown in Fig. 2 (b).

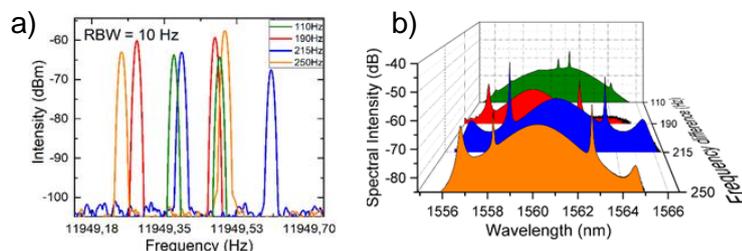

Fig. 2. a) RF spectra Tunability of the dual-comb frequency difference b) OSA spectra. The four plots correspond to the following frequency differences, starting from the front to the back 250Hz, 215 Hz, 190 Hz, 110 Hz.

Figure 3 presents the results for dual-comb temporal stability, namely for dual-comb operation with a repetition rate difference of 205 Hz. The dual-comb was monitored for 400 minutes in room temperature conditions. The regime tends to remain the same after that time, with a slight shift in repetition rate and central wavelength. Even though there is drift over time in the carrier-envelope offset frequency, the offset between the two combs shows stability with a drift of 6 Hz over 6 hours, going from 205 Hz to 199 Hz unevenly without any additional external stabilization (Fig. 3 a). The resolution bandwidth (RBW) of the RF spectrum analyzer was 1 Hz. As follows from the Fig. 3 (a), the short-term stability over 1 min can be estimated as 15 mHz. The drift is more than twice less than the drift of 38 mHz/80 sec obtained in [12].

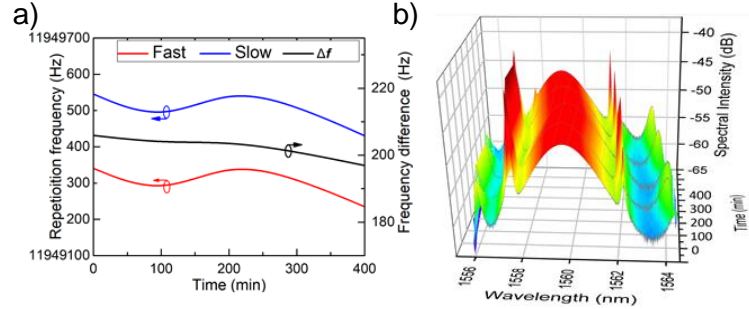

Fig. 3. Stability of the dual-comb (a) frequencies of fast and slow axes and their difference; (b) optical spectra over time.

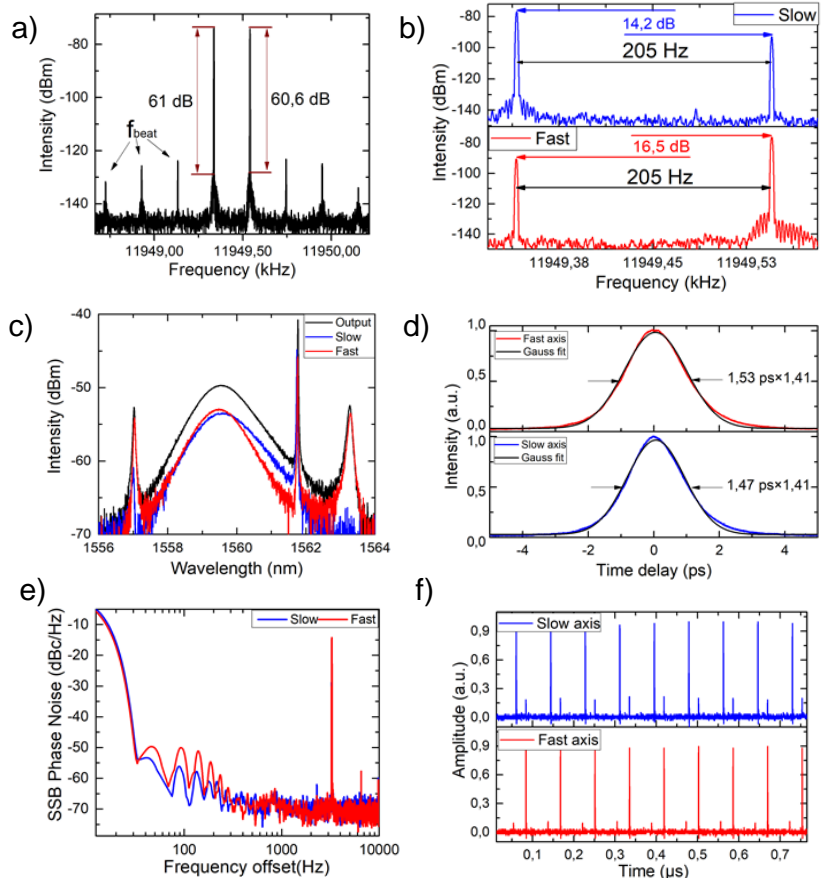

Fig. 4. (a) RF spectrum of the original signal with frequency difference 205 Hz, (b) separated signal by PBS, in red color the fast axis, in blue color the slow axis, (c) optical spectra, (d) autocorrelation traces, (e) RF phase noise, (f) oscilloscope traces for each comb after PBS.

A polarization beam splitter (PBS1550SM-APC Thorlabs), in combination with a polarization controller (PC), was employed at the laser output to separate the signal into two linear orthogonal states of polarization. The PBS has an excitation ratio (splitter) ≥ 20 dB at an operating wavelength of 1550 ±40 nm. Fig. 4 (a) shows the RF spectra of the common signal from the output port before PBS. Two frequency components with a 205 Hz repetition rate difference were fixed at 11.949340 and 11.949545 MHz. From these frequencies we obtain a refractive index difference of $2.573 \times 10^{-5}$, a beat length of 0.0605 meters, a difference in propagation between both pulses of 103.925 $m^{-1}$ and a phase delay of 1700.52. Signal-to-noise ratios of more than 60 dB are observed for both RF peaks, which is evidence of good stability of the dual combs. These two peaks have FWHM bandwidths of 1.4 Hz and 1.5 Hz. Between the two combs, the weaker frequency peaks are beat notes ($f_{beat}$), equally spaced by the comb's frequency difference. A beat notes has a FWHM bandwidth of 1.5 Hz. Fig. 4 (b) depicts the RF spectra of the two orthogonally polarized outputs at RBW of 1 Hz. The maximum difference in amplitude of around 15 dBm between the components was achieved by adjusting the PC before the PBS. In other words, two pulse trains are in a state of polarization that is close to orthogonal. Fig. 4 (c) illustrates the common optical spectrum of polarization-multiplexed mode-locking and spectra for each comb individually. In the slow axis, the center wavelength is 1559.45 nm with Δλ is 1.85 nm and in the fast axis, the center wavelength is 1559.62 nm with Δλ is 1.91 nm.

The optical spectra of the output, fast and slow axis, are overlapped with just a small swift of the fast axis towards the left (smaller frequencies). Although the PBS can separate the two combs to a certain level of the extinction ratio of 20 dB, the observed comb separation with a lower extinction ratio can be caused by imperfections in the adjustment of the output polarization controller to PBS. Autocorrelation traces (AC) for both pulse trains are also displayed in Fig. 4 (c). The pulse duration delivered by this laser was measured at its full width at half maximum (FWHM) for both axes using an autocorrelator (Femtochrome FR-103XL) and the results were 1.53 ps for the fast-moving pulse and 1.47 ps for the slow-moving pulse. Also, Fig. 4 (d) demonstrates both pulses have a shape close to Gaussian. When it comes to the power the overall power output of the laser is 0.05 mW.

The study of the polarization dynamics of the fast and slow axis in comparison with the dual-comb signal has been performed with a fast polarimeter (Novoptel PM1000-XL-FA-N20-D) with a resolution of 10 ns. Polarization data were acquired for every sample with different frequency differences. Each trace comprises samples of the Stokes parameters $S_0$, $S_1$, $S_2$, $S_3$, and degree of polarization (DOP) DOP $= \sqrt{\langle S_1 \rangle^2 + \langle S_2 \rangle^2 + \langle S_3 \rangle^2}/\langle S_0 \rangle$. Where <...> means averaging over the 320 ns. Figure 5 (a-c) shows the normalized Stokes parameters $s_i = \langle S_i \rangle/\sqrt{\langle S_1 \rangle^2 + \langle S_2 \rangle^2 + \langle S_3 \rangle^2}$, the output power $S_0$, DOP (a, b) and SOP evolution in the Poincaré sphere (c) as a function of time. Given the different overlapping of the polarization multiplexed combs as a function of time, the normalized Stokes parameters, the output power and DOP the maximum demonstrate the oscillatory behavior in the time domain as shown in Fig. 5 (a, b) with an asymptotic value corresponding the absence of the overlapping. As a result of the oscillation of the Stokes parameters, the trajectory at the Poincare sphere takes the form of arc as shown in Fig. 5 c. The oscillatory behavior of Stokes parameters can be used for evolution of the polarimetric signatures of light scattered by the object in terms of Mueller matrix [21]. Instead of generating four different input SOPs, analysis of Fourier components of the Stokes parameters oscillations for scattered light vs input Stokes parameters enables extraction of the Mueller matrix elements. When the polarization dynamics were measured after the comb separation using a PBS, the degree of polarization of the fast and slow axes were 0.98 and 0.99 respectively for the optimal separation level.

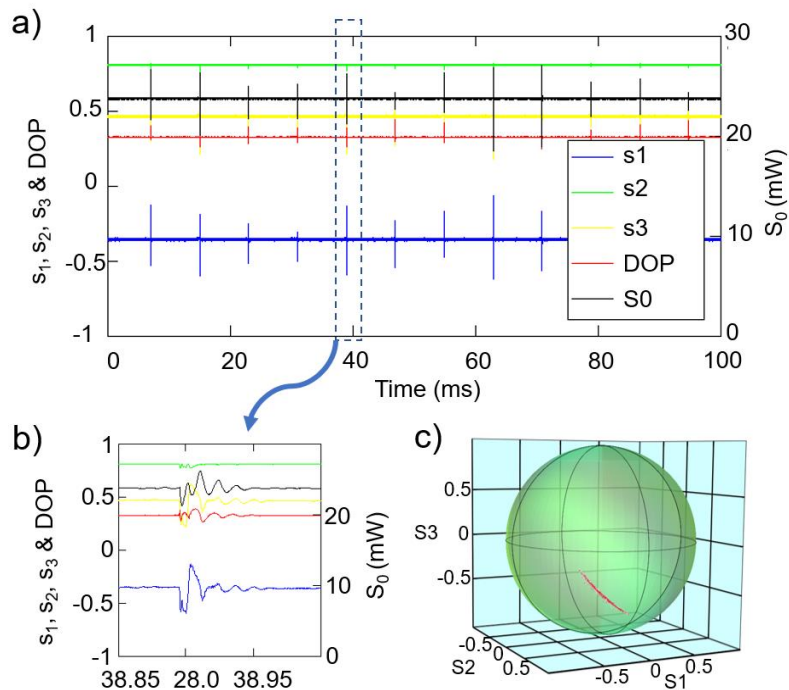

Fig. 5. The normalized Stokes parameters $s_i$ (i=1,2,3), the output power $S_o$, DOP (a, b) and SOP evolution in the Poincaré sphere (c) as a function of time.

In conclusion, we designed and characterized experimentally the polarization-multiplexed system capable of generating two optical frequency combs with repetition rate difference in the range of a few hundred Hz. These dual-comb regimes are relatively stable, having a drift of 6 Hz over 6 hours (or 15 mHz in 60 sec) without thermo stabilization. The analysis of the polarization dynamics and the combs' separation using a PBS suggest that the imperfections in the extinction ratio of polarization beam splitter and adjustment of SOP at the input of PBS can lead to the demonstrated extinction ratio for two combs of 16.5 dBm. The obtained frequency difference stability and analyses of polarization dynamics conclude that this polarization-multiplexed dual-comb source can be used to design the polarimetric LIDAR able to recognize the object's texture based on the polarimetric signatures.

**Disclosures.** The authors declare no conflicts of interest.

**Data availability.** Data underlying the results are not publicly available but may be obtained from the authors upon request.